\newcommand{\acs}{}
\newcommand{\gls}{}
\newcommand{\acp}{}

\newcommand{\acrfullpl}{}

\newcommand{\fermilat}{\textit{Fermi}-LAT}

\documentclass{PoS}
\RequirePackage{nccmath}
\RequirePackage{enumerate} 
\RequirePackage{subcaption}
\RequirePackage{wrapfig} 

\title{Very high energy gamma-ray follow-up observations of novae and dwarf novae with the MAGIC telescopes }

\ShortTitle{Novae MAGIC}

\author{\speaker{Rub\'en L\'opez-Coto}\\
        IFAE, Campus UAB, E-08193 Bellaterra, Spain\\
	E-mail: \email{rlopez@ifae.es}
        }

\author{Julian Sitarek, Wlodek Bednarek\\
        University of \L\'od\'z, PL-90236 Lodz, Poland\\
}        

\author{E.~de O\~na Wilhelmi for the MAGIC Collaboration\\
Institute of Space Sciences, E-08193 Barcelona, Spain\\
}

\author{R. Desiante\\
Universit\`a di Udine, and INFN Trieste, I-33100 Udine, Italy\\
}

\author{F. Longo\\
Universit\`a di Trieste and INFN Trieste, Italy\\
}

\author{E. Hays for the \fermilat\ Collaboration\\
NASA Goddard Space Flight Center, Greenbelt, MD 20771, USA\\
}

\abstract{In the last few years the Fermi-LAT instrument has detected GeV gamma-ray emission from several novae. Such GeV emission can be interpreted in terms of inverse Compton emission from electrons accelerated in the shock or in terms of emission from hadrons accelerated in the same conditions. The latter might reach much higher energies and could produce a second component in the gamma-ray spectrum at TeV energies. We perform follow-up observations of selected novae and dwarf novae in search of the second component in TeV energy gamma rays. This can shed light on the acceleration process of leptons and hadrons in nova explosions. We have performed observations with the MAGIC telescopes of 3 sources, a symbiotic nova YY Her, a dwarf nova ASASSN-13ax and a classical nova V339 Del, shortly after their outbursts. We did not detect TeV gamma-ray emission from any of the objects observed. The TeV upper limits from MAGIC observations and the GeV detection by Fermi constrain the acceleration parameters for electrons and hadrons.
}

\FullConference{The 34th International Cosmic Ray Conference,\\
		30 July- 6 August, 2015\\
		The Hague, The Netherlands}

\begin{document}


\section{Introduction}
A classical nova is a thermonuclear runaway leading to the explosive ejection of the envelope accreted onto a white dwarf (WD) in a binary system in which the companion is either filling or nearly filling its Roche surface \cite{classicalnovae}.
They are a type of cataclysmic variables, i.e. optically variable binary systems with a mass transfer from a companion star to WD.
Novae are typically detected first in optical observations when the brightness of the object increases by 7-16 magnitudes before fading back to the original level. 
The energy spectra of novae often contain a thermal X-ray continuum. 
The symbiotic novae, like the classical novae, are also initiated by a thermonuclear explosion on the surface of the WD.
However in the case of symbiotic novae, the WD is deep immersed in the wind of a late-type companion star (see e.g. \cite{sh12}).

The diffusive shock acceleration at the blast wave of symbiotic novae was expected to accelerate particles up to energies of a few TeVs \cite{th07}.
In 2010 the first GeV $\gamma$-ray emission was detected by \textit{Fermi}-\acs{LAT} from the symbiotic nova V407 Cyg \cite{ScienceFermi}.
In this contribution we present the results of the follow-up observations of three selected novae and dwarf novae: the symbiotic nova YY Her, the dwarf nova ASASSN-13ax and the classical nova V339 Del.

The $\gamma$-ray emission has been subsequently explained in terms of leptonic or hadronic models \cite{ScienceFermi,novaescience}. 
Local radiation fields create a target for the \acs{IC} scattering of the electrons.
Protons accelerated in the same conditions can interact with matter producing gamma rays via proton-proton interactions.
For instance, \cite{Sitarek} attribute the GeV $\gamma$-ray emission to the \acs{IC} process on the strong radiation field of the red giant.
The same model predicts a second component in the TeV range due to proton-proton interactions with the wind of the red giant.
Also \cite{md13} consider acceleration of leptons and hadrons in the nova shock.
In that model the magnetic field, which determines the acceleration efficiency, is obtained assuming equipartition with the thermal energy density upstream of the shock.
The GeV $\gamma$-ray emission is then a product of \acs{IC} scattering of the nova light by the electrons.

In the last few years \textit{Fermi}-\acs{LAT} has discovered GeV $\gamma$-ray emission from a few more novae: V1324 Sco, V959 Mon, V339 Del, V1369 Cen, V745 Sco and the very recent Nova Sagittarii 2015 \cite{ATELCen2013,ATELSco2014, novaescience, ch15}. 
Most of these sources are \acp{CN}. 
Contrary to the symbiotic ones, the wind of the companion star is not strong, but they all show similar spectral properties. 
In \acp{CN} the particle acceleration can occur e.g. on a bow shock between the nova ejecta and the interstellar medium or in weaker internal shocks due to inhomogeneity of the nova ejecta \cite{novaescience}.
In particular, orbital motion of the system can shape the nova ejecta into a faster polar wind and a denser material expanding on the equatorial plane \cite{cho14}.

So far no VHE $\gamma$-ray emission has been detected from any nova event. 
\acs{VERITAS} observations of V407 Cyg starting 10 days after the nova explosion yielded a differential \acs{U.L.} on the flux of $2.3 \times 10^{-12}\, \mathrm{erg\,cm^{-2}\,s^{-1}}$ at 1.6 TeV \cite{al12}.

\section{Instruments} \label{sec:ins}
The three sources were first detected and observed by optical instruments. 
The results of the MAGIC observations were supported by the analysis of quasi-simultaneous \fermilat\ observations. 

\subsection{MAGIC telescopes}
The VHE gamma-ray observations were performed with the MAGIC telescopes. 
MAGIC is a system of two 17\,m Cherenkov telescopes located on the Canary Island of La Palma at a height of 2200 m a.s.l.
The telescopes can perform observations of gamma rays with energies as low as $\sim$50\,GeV and up to tens of TeV. 
During Summer 2011 and 2012 MAGIC underwent a major upgrade. 
After the upgrade the sensitivity of the MAGIC telescopes in the best energy range ($\gtrsim300\,$GeV) is $\sim 0.6\%$ of Crab Nebula flux in 50\,h of observations \cite{Performance2013}.
All the data used for this paper were taken after the upgrade.
The data were analyzed using the standard analysis chain \cite{MARS, Performance2013}.
The significance of a gamma-ray excess was computed according to Eq.~17 of \cite{LiMa}.
The upper limits on the flux were calculated following the approach of \cite{Rolke} using 95\% C.L. and accounting for a possible 30\% systematic uncertainty on the effective area of the instrument. 

\subsection{\fermilat}
\fermilat\ is a pair-conversion telescope launched in 2008 that detects photons with energies from $20\,$MeV to $>300$\,GeV \cite{at09}.
Thanks to a large \acs{FoV} ($\sim 2.4$ sr), the \fermilat\ observatory, operated in scanning mode, provides coverage of the full sky every three hours enabling searches for transient sources and overlap with ground-based observatories.
The \acs{LAT} data were analyzed in the energy range 100 MeV $-$ 300 GeV using an unbinned maximum likelihood method as implemented in the {\it Fermi} Science Tools v9r32p5, the P7REP\_SOURCE\_V15 LAT \acrfullpl{IRF}, and associated standard Galactic and isotropic diffuse emission models\footnote{The P7REP data, IRFs, and diffuse models (gll\_iem\_v05.fit and iso\_source\_v05.txt) are available at http://fermi.gsfc.nasa.gov/ssc.}.
We selected events within a \acs{RoI} of $15^\circ$ centered on the \acs{LAT} best position reported by \cite{novaescience} for V339 Del and required a maximum \acs{Zd} of $100^\circ$ in order to avoid contamination from Earth limb photons. Additionally, we applied a gtmktime filter (no.3) recommended for combined survey and pointed mode observations\footnote{http://fermi.gsfc.nasa.gov/ssc/data/analysis/documentation/Cicerone/\\Cicerone\_Likelihood/Exposure.html}, selecting good quality data at times when either the rocking angle was less than $52^\circ$ or the edge of the analysis region did not exceed the maximum \acs{Zd} at $100^\circ$.
Sources from the \acs{2FGL} catalogue located within $20^\circ$ the \acs{RoI} were included in the model used to perform the fitting procedure. The \acp{U.L.} were calculated at 95\% \acs{C.L.} using the Bayesian method provided with the \textit{Fermi} Science Tools\footnote{http://fermi.gsfc.nasa.gov/ssc/data/analysis/scitools/python\_tutorial.html}.

\section{Results}

\subsection{YY Her}\label{sec:yyher}

\begin{minipage}[h]{0.45\textwidth}
\vspace{0.3 cm}

YY Her is a symbiotic nova system that undergoes a recurrent pattern of outbursts. 
MAGIC observations of YY Her occurred on \acs{MJD} 56404 and could not be continued due to the Full Moon.  
No significant VHE gamma-ray emission was detected. 
We computed flux upper limits at 95\% confidence level obtaining $<5.0\times \mathrm{10^{-12} ph\,cm^{-2}\,s^{-1}}$ above 300 GeV. 
Also in \fermilat\ no emission was detected over a longer interval from MJD 56392.5 to 56412.5. 
Upper limits at 95\% confidence level were set as $2.8 \times 10^{-8} \mathrm{ph\,cm^{-2}\,s^{-1}}$ above 100\,MeV.
Differential upper limits obtained from the \fermilat\ and MAGIC observations of YY Her are shown in Fig.~\ref{fig:yyher_ul}.

\vspace{0.5 cm}
\end{minipage}
 \hspace{0.01\textwidth}
\begin{minipage}[h]{0.48\textwidth}
\centering
\includegraphics[width=\textwidth]{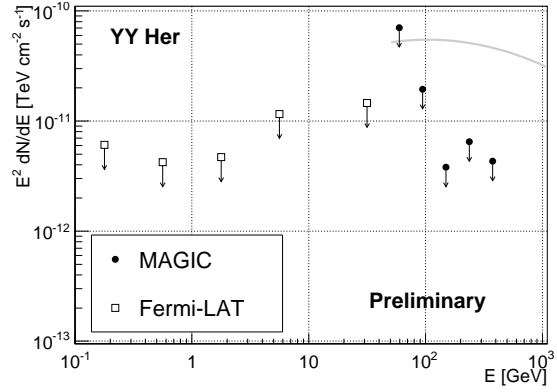}
\captionof{figure}{Differential upper limits on the flux from YY Her as measured by the \fermilat\ (empty squares) and MAGIC (full circles). 
See text for details of the time ranges covered by the points.
For comparison a spectrum of Crab Nebula is shown with a gray curve. 
}\label{fig:yyher_ul}

\end{minipage}

\subsection{ASASSN-13ax}\label{sec:asassn}
ASASSN-13ax is a member of a different class of cataclysmic variables, the dwarf novae, which are known for significantly weaker optical outbursts (2-6 magnitudes) than classical novae. 

\begin{minipage}[h]{0.48\textwidth}
\centering
\vspace{0.3 cm}
\includegraphics[width=\textwidth]{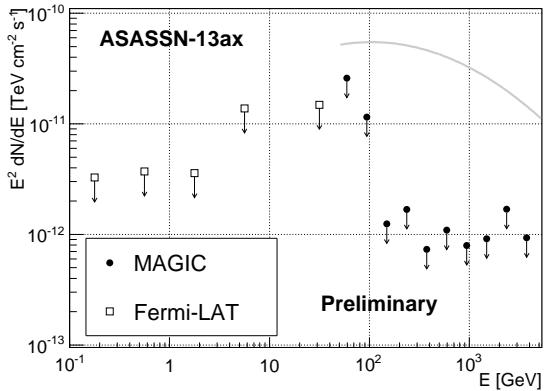}
\captionof{figure}{Differential upper limits on the flux from ASASSN-13ax as measured by the \fermilat\ (empty squares) and MAGIC (full circles). 
See text for details of the time ranges covered by the points.
For comparison a spectrum of Crab Nebula is shown with a gray curve. 
}\label{fig:asassn_ul}

\vspace{0.5 cm}
\end{minipage}
 \hspace{0.01\textwidth}
\begin{minipage}[h]{0.45\textwidth}

Instead of undergoing a thermonuclear explosion on the surface of the WD, these outbursts are caused by the gravitational energy release from a partial collapse of the accretion disk surrounding the WD.
The MAGIC observations were performed on two consecutive nights starting \acs{MJD} 56478. 
In the absence of detectable VHE emission, upper limits at 95\% confidence level were set as $<1.5\times \mathrm{10^{-12} ph\,cm^{-2}\,s^{-1}}$ above 300 GeV.
 \emph{Fermi}-\acs{LAT} observations put a 95\% \acs{C.L.} \acs{U.L.} on the flux of the source above 100\,MeV at the level of $1.6 \times 10^{-8} \mathrm{cm^{-2}\,s^{-1}}$ in the time period \acs{MJD} 56468.5 to 56488.5.
Differential upper limits obtained from the \fermilat\ and MAGIC observations of ASASSN-13ax are shown in Fig.~\ref{fig:asassn_ul}.

\end{minipage}

\subsection{V339 Del}\label{sec:v339del}

\begin{minipage}[h]{0.45\textwidth}
\vspace{0.3 cm}

V339 Del was a fast, classical nova detected by optical observations on observations on \acs{MJD} 56520 (CBET \#3628). 
The nova was exceptionally bright reaching a magnitude of V$\sim 5\,$mag (see top panel of Fig.~\ref{fig:del_mwl}), and it triggered follow-up observations at frequencies ranging from radio to VHE gamma-rays.
Photometric measurements suggest a distance for V339 Del of $4.5\pm0.6$\,kpc \cite{sch14}.
The spectroscopic observations performed on MJD 56522.1 revealed emission wings extending to about $\pm 2000\,$km/s and a Balmer absorption component at a velocity of $600\pm 50\,$km/s \cite{sh13}.
The pre-outburst optical images revealed the progenitor of nova V339 Del to be a blue star \cite{de13}. 
Originally MAGIC observations of V339 Del were motivated by its extreme optical outburst.
Soon after MAGIC started observations they were additionally supported by the detection of GeV emission by the \fermilat\ from the direction of V339 Del.
The MAGIC observations started already on the night  of \acs{MJD} 56520, however they were marred by bad weather conditions. 
The good quality data used for most of the analysis span 8 nights between \acs{MJD} 56529 and 56537. 
The total effective time was 11.6\,h.
In addition to the nightly upper limits we performed a dedicated analysis of the poor quality (affected by calima, a dust layer originating from Sahara) night of \acs{MJD} 56520.
We applied an estimated energy and collection area corrections based on LIDAR measurements \cite{Lidar_icrc}.
No VHE gamma-ray signal was found from the direction of V339 Del.

\vspace{0.5 cm}
\end{minipage}
 \hspace{0.01\textwidth}
\begin{minipage}[h]{0.48\textwidth}
\centering
\includegraphics[width=\textwidth]{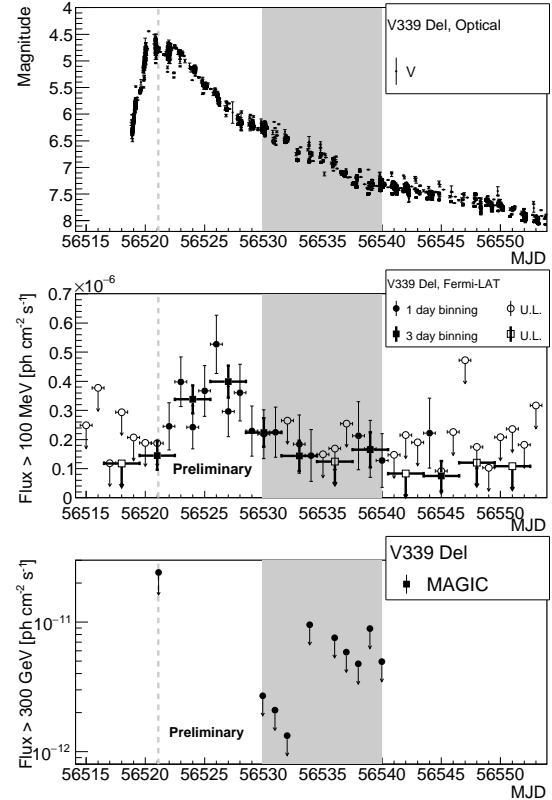}
\captionof{figure}{
Multiwavelength light curve of V339 Del during the outburst in August 2013.
Top panel: optical observations in the V band obtained from AAVSO-LCG\protect\footnote{http://www.aavso.org/lcg} service.
Middle panel: the \fermilat\ flux (filled symbols) and upper limits (empty symbols) above 100 MeV in 1-day (circles, thin lines) or 3-day (squares, thick lines).
A 95\% C.L. flux upper limit is shown for time bins with TS$<$4.
Bottom panel: Upper limit on the flux above 300 GeV observed with MAGIC telescopes.
The gray band shows the observation nights with MAGIC. 
The dashed gray line shows a MAGIC observation night affected by bad weather.}
\label{fig:del_mwl}

\end{minipage}

\begin{figure}
\centering
\includegraphics[width=0.49\textwidth]{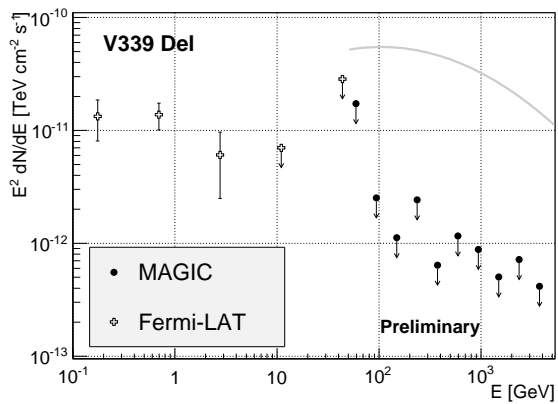}
\caption{
Differential upper limits on the flux from V339 Del as measured by MAGIC (filled circles) and the flux measured by \fermilat\ (empty crosses) in the same time period, MJD 56529 to 56539.
For comparison a spectrum of Crab Nebula is shown with a gray curve. 
}\label{fig:del2013_sed}
\end{figure}

We computed a night by night integral upper limit above 300\,GeV (see bottom panel of Fig.~\ref{fig:del_mwl}. 
The differential upper limits for the whole good quality data set computed in bins of energy are shown in Fig.~\ref{fig:del2013_sed}.

Nova V339 Del was the subject of a \fermilat\ \gls{ToO} observation \cite{AT5302} triggered by the optical discovery (CBET \#3628); the \acs{ToO} started on \acs{MJD} 56520 and lasted for six days.
The $\gamma$-ray emission from V339 Del was first detected by \fermilat\ in 1-day bins on \acs{MJD} 56522 \cite{novaescience}. 
The emission peaked on \acs{MJD} 56526 and entered a slow decay phase afterwards (Figure~\ref{fig:del_mwl}).
For the light curves, the data were fitted using a power-law spectral model initially leaving the spectral index and the normalization free to vary.
We then fixed the spectral index at the average value of 2.3 calculated over the most significant detections (\acs{TS}$>$9) to generate the plots shown in the middle panel of Figure~\ref{fig:del_mwl}. 
The \acs{LAT} \acs{SED} of V339 Del shown in Figure~\ref{fig:del2013_sed} was extracted in five logarithmically spaced energy bins from 100 MeV to 100 GeV.
Similarly to the light curves, energy binned data shown in Figure~\ref{fig:del2013_sed} were fitted using a simple power-law and showing a 95\% \acs{C.L.} \acp{U.L.} for \acs{TS}$<$9. 
In the period coincident with the MAGIC observations (\acs{MJD} 56529 to 56539) the \fermilat\ spectrum can be described by an effective power-law with a spectral index of $2.37\pm0.17$ and flux above 100 MeV of $(0.15\pm 0.04) \times 10^{-6} \mathrm{cm^{-2}\,s^{-1}}$. 
The rather low statistical significance (\acs{TS}=49) does not constrain the value of an exponential cut-off of the emission in this period.
Note, however, that the most energetic photon, with $E=5.9$\,GeV was recorded on \acs{MJD} 56534, i.e. within the time period covered by MAGIC.
The \fermilat\ analysis for a broader time range, \acs{MJD} 56526 to 56547, covering the whole decay phase of the \fermilat\ light curve allowed us to obtain a more significant signal with a \acs{TS} of 121. 
Nevertheless  we obtain a similar value of flux above 100 MeV,  $(0.13\pm 0.03) \times 10^{-6} \mathrm{cm^{-2}\,s^{-1}}$, for this broader period.
The spectrum in this case can be described, with a $3.3\sigma$ significance higher with respect to the simple power-law, by an exponentially cut-off power-law with a spectral index of $1.44\pm 0.29$ and a cut-off energy of $1.6\pm0.8$\,GeV. 

\section{Conclusions}\label{sec:conc}
The MAGIC telescopes performed observations of 3 objects: the symbiotic nova YY Her, the dwarf nova ASASSN-13ax and the classical nova V339 Del. 
No significant VHE gamma-ray emission was found from the direction of any of them.
Out of these three objects, V339 Del is the only one detected at GeV energies.
It has also extensive optical observations which shed some light on both the companion star and the photosphere of the nova.
Therefore it has the highest potential for constraining the leptonic and hadronic processes in novae.
MAGIC will continue follow-up observations of the promising novae candidates in the following years.

\bigskip 
\begin{acknowledgments}
The MAGIC Collaboration would like to thank
the Instituto de Astrof\'{\i}sica de Canarias
for the excellent working conditions
at the Observatorio del Roque de los Muchachos in La Palma.
The financial support of the German BMBF and MPG,
the Italian INFN and INAF,
the Swiss National Fund SNF,
the ERDF under the Spanish MINECO (FPA2012-39502), and
the Japanese JSPS and MEXT
is gratefully acknowledged.
This work was also supported
by the Centro de Excelencia Severo Ochoa SEV-2012-0234, CPAN CSD2007-00042, and MultiDark CSD2009-00064 projects of the Spanish Consolider-Ingenio 2010 programme,
by grant 268740 of the Academy of Finland,
by the Croatian Science Foundation (HrZZ) Project 09/176 and the University of Rijeka Project 13.12.1.3.02,
by the DFG Collaborative Research Centers SFB823/C4 and SFB876/C3,
and by the Polish MNiSzW grant 745/N-HESS-MAGIC/2010/0.
JS is supported by Fundacja U\L. 

The \textit{Fermi} LAT Collaboration acknowledges generous ongoing support
from a number of agencies and institutes that have supported both the
development and the operation of the LAT as well as scientific data analysis.
These include the National Aeronautics and Space Administration and the
Department of Energy in the United States, the Commissariat \`a l'Energie Atomique
and the Centre National de la Recherche Scientifique / Institut National de Physique
Nucl\'eaire et de Physique des Particules in France, the Agenzia Spaziale Italiana
and the Istituto Nazionale di Fisica Nucleare in Italy, the Ministry of Education,
Culture, Sports, Science and Technology (MEXT), High Energy Accelerator Research
Organization (KEK) and Japan Aerospace Exploration Agency (JAXA) in Japan, and
the K.~A.~Wallenberg Foundation, the Swedish Research Council and the
Swedish National Space Board in Sweden.
Additional support for science analysis during the operations phase is gratefully acknowledged
from the Istituto Nazionale di Astrofisica in Italy and the Centre National d'\'Etudes Spatiales in France.

\end{acknowledgments}

\bigskip 

\bibliographystyle{JHEP}
\bibliography{references} 

\end{document}